\documentclass{osa-article}

\journal{osajournal}

\articletype{Research Article}

\begin{document}

\newcommand{\edited}[1]{\textcolor{black}{#1}}

\title{Fundamental limit of single-mode integral-field spectroscopy.}

\author{S. Y. Haffert \authormark{1,2}}

\address{\authormark{1}Steward Observatory, University of Arizona}
\address{\authormark{2}NASA Hubble Fellow}
\email{\authormark{1}shaffert@arizona.edu} 



\begin{abstract}
There are several high-performance adaptive optics systems that deliver diffraction-limited imaging on ground-based telescopes, which renewed the interest of single-mode fiber (SMF) spectroscopy for exoplanet characterization. However, the fundamental mode of a telescope is not well matched to those of conventional SMFs. With the recent progress in asphere manufacturing techniques it may be possible to reshape the fundamental mode of a SMF into any arbitrary distribution. An optimization problem is setup to investigate what the optimal mode field distribution is and what the fundamental throughput limit is for SMF spectroscopy. Both single-object spectrographs and integral-field spectrographs are investigated. The optimal mode for single-object spectrographs is found to be the aperture function of the exit pupil, while for integral-field spectrographs the optimal mode depends on the spatial sampling of the focal plane. For dense sampling, a uniform mode is optimal, while for sparse sampling, the mode of a conventional SMF is near optimal. With the optimal fiber mode, high throughput (>80\,\%) can be achieved when the focal plane is (super) Nyquist sampled. For the Nyquist sampled cases, the optimal mode has almost 20\,\% more throughput than a conventional SMF.
\end{abstract}

\section{Introduction}
The discovery of the first exoplanet in 1995 \cite{mayor1995jupiter} led to a rapid increase of research into exoplanets, which was driven by an ever increasing amount of dedicated instruments and exoplanet observing techniques. The most successful observing technique uses the transit method, where an exoplanet moves in front of its host star and the light from the star passes through the planets atmosphere. This leaves a small imprint of the planets atmosphere in the observed spectrum, which can be used to characterize the exoplanet. However, because the planet and star are usually unresolved, starlight is the major contribution of noise. If we want to probe exoplanets at higher contrast, the influence of the star has to be decreased. Spatially resolving the host star from planet can reduced this influence by many orders of magnitude.


Direct imaging instruments \cite{macintosh2014first, beuzit2019sphere, males2020magao} are specifically designed to spatially resolve planets, that are on a large enough orbit, from their host stars. On ground-based telescopes, these instruments need to use Extreme Adaptive Optics (ExAO) systems to remove the effects of atmospheric turbulence to restore the telescopes spatial resolution. Advanced coronagraphs are then used in conjunction with the ExAO system to suppress the starlight, which enhances the contrast between the star and planet even further \cite{guyon2018extreme}. After employing all these optical techniques, there is still residual starlight that leaks through the system due to imperfect AO correction and residual phase aberrations that creates speckles which can appear as planets. Image processing techniques are necessary to differentiate these residual speckles from the planet signal. A very powerful post-processing technique is high-resolution spectroscopy (HRS), which uses the difference in spectral lines between the planet and star to differentiate the detected signals \cite{sparks2002imaging, snellen2015combining}. 

Many instruments are currently under development that combine high-contrast imaging with high-resolution spectroscopy. Single-mode fibers are a promising approach to coupling direct imaging instruments with spectrographs \cite{jovanovic2017efficient}. The advantage of SMFs is that they are diffraction-limited, which leads to small and stable spectrographs \cite{bland2006instruments, bland2010pimms}. This is a strong advantage for accurate post-processing because the spectral stability determines how well the residual starlight can be disentangled from the planet light. \edited{The stability of the line spread function is also the reason why SMF spectrographs are gaining traction as the fiber link for high-precision Radial Velocity machines such as iLocator \cite{crepp2016ilocater}.} Additionally, the mode filtering capabilities of SMF reject residual wavefront errors from the AO system \cite{mawet2017observing}. And finally, SMF can be used to design more efficient coronagraphs \cite{ruane2018efficient, por2020single, haffert2020single}. Several ExAO systems have such a capability installed \cite{mawet2016keck, anagnos2020innovative} or planned \cite{vigan2018bringing, haffert2020diffraction}.

A major problem for SMF spectrographs is that the fundamental mode is not matched to the mode of the telescope, which limits the throughput to 82 percent \cite{shaklan1988coupling}. For single-object spectrographs this is not a major problem, because only the on-axis object needs to be injected. For such systems the coupling efficiency can be improved with multi-element aspheric optics that reshape the incoming wavefront. The Phase Induced Amplitude Apodization (PIAA) optics were used at SCExAO for this exact purpose \cite{jovanovic2017efficient}. PIAA optics \edited{are a set of freeform optics placed close to or in the pupil that} reshape the incoming wavefront \edited{from the telescope} to match the fundamental mode of the fiber without any loss, except for the Fresnel losses. \edited{A similar approach is envisioned for the SMF injection unit of KPIC \cite{calvin2021enhancing}.} For integral-field spectroscopy the situation is not as simple. SMF integral-field spectrographs use micro-lenses to increase the fill-factor. Each microlens injects light into an independent fiber. This works great for objects that are on-axis with respect to any of the microlens centers. However, offsets lead to spatial variations in the coupling efficiencies and can even create dead-zones where there is no throughput at all \cite{corbett2009sampling, por2020single}.

Additive manufacturing of micro-optics has made giants steps in the past years. Currently multi-element aspheric micron scaled optics can be created \edited{and written directly on top of fibers} \cite{dietrich2018situ}. The technology has been used to create micro-lens arrays that feed compact multi-core fibers that are used as input to spectrographs \cite{anagnos2020innovative, haffert2020diffraction}. However, the micro-lens designs to date, consist only of single-element spherical singlets. \edited{It may be possible to shape the fundamental mode of the fiber with multi-element freeform micro-lens arrays and increase the throughput.} Improved mode-coupling may also lead to smaller spatial variations. \edited{Freeform micro-optics directly written on top of a SMF that reshape the mode to a tophat or a donut-like distribution have already been created \cite{gissibl2016sub} and creating an array of such beamshaping optics could feed a SMF array (or a single Multi Core Fiber). The manufacturing technology for arbitrary mode shaping is available, however it is not clear how much can be gained if such a technology is used.} In this paper I derive the equations that can determine the fundamental coupling limit for SMF integral-field spectroscopy. The approach that is used in this paper is independent of the exact fundamental mode of the fiber and the \edited{method to reshape the beam}, and therefore provides a general limit. In Section 2, I set up the theoretical optimization problem and in section 3 I compare the throughput between conventional fibers and optimal fibers. Section 4 concludes the paper with an outlook.

\section{Theoretical model}
\subsection{Setting up the optimization problem for a single fiber}
\begin{figure}
\includegraphics[width=\textwidth]{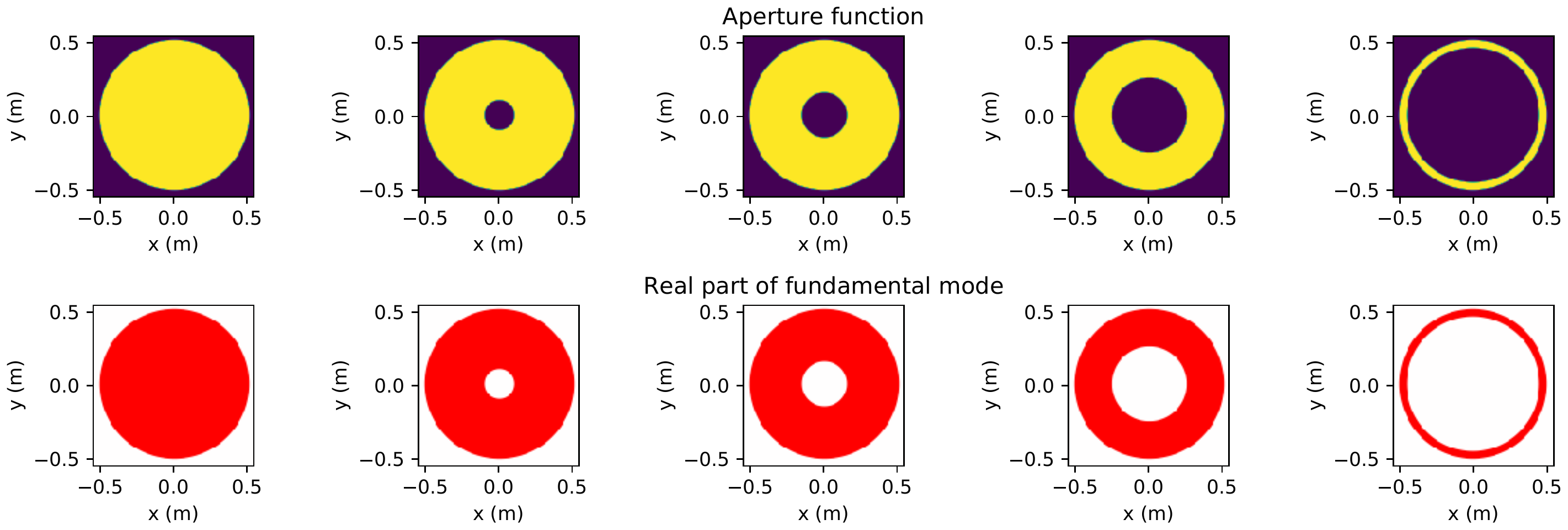}
\caption{The fundamental mode for (un)obstructed apertures. The modes are symmetric and completely real. The fundamental mode is the aperture function itself for any aperture shape. }
\label{fig:optimal_on_axis}
\end{figure}

\edited{Any scalar electric field $E$ can be expanded onto an orthogonal mode basis with modes $\{\phi_n\}$ and mode coefficients $\{\alpha_n\}$ as,
\begin{equation}
    E = \sum_n \alpha_n \phi_n.
\end{equation}
The plane wave expansion is an example of such a mode basis. The coefficient of a particular mode can be found by calculating the overlap integral between the electric field and the corresponding mode field distribution,
\begin{equation}
    \alpha_n = \int_A E \phi_n^*dA.
\end{equation}
The integral is taken over the area $A$ that contains both electric fields.} We assume that the modes have been normalized \edited{in such a way that $\int_A |\phi_n|^2 dA=1$}, for the ease of calculations. The coefficient \edited{$\alpha_n$} is an electric field coefficient and is therefore a complex number. \edited{From here on, the subscript $n$ is dropped because only a single mode is relevant for single-mode fibers.} Due to the linearity of Maxwell's Equations, the problem can always be setup in such a way that the \edited{modal coupling is calculated in the pupil by back propagating the fiber output through the optical system to the pupil plane}. This does assume that the optical system is lossless, which means there is no loss of information during the forward or backward propagation. A general throughput factor does not matter. In \edited{this} section, both $E$ and $\phi$ are defined in the pupil. The power that is \edited{coupled} into mode $\phi$ can be found by simply taking the absolute value squared of $\alpha$,
\begin{equation}
    I = |\alpha|^2 = \alpha \cdot \alpha^H
\end{equation}
Here $\alpha^H$ denotes the Hermitian transpose of $\alpha$. This equation can be used to determine the coupling efficiency,
\begin{equation}
    \eta = \frac{I}{I_{\mathrm{in}}} = \frac{|\alpha|^2}{\int_A |E|^2 dA}.
\end{equation}
With $I_{\mathrm{in}}$ the input power of the electric field and $\eta$ the coupling efficiency. For numerical optimization, the electric field and the fiber mode are discretized, which leads to the following vector notation,
\begin{equation}
    \alpha = \sum_i w_i E_i \phi_{i}^* = \vec{\phi}^H W \vec{E}.
\end{equation}
The subscript \edited{$i$} indicates the field value at discrete element $i$ with $w_i$ as the weight of that element. The sum can also be rewritten in a weighted dot product with $W$ the weight matrix. The weight matrix is a diagonal matrix created from the weight vector. Substituting this equation in the power equation gives,
\begin{equation}
    I_n = \alpha \cdot \alpha^H = \vec{\phi}^H W \vec{E}\vec{E}^H W \vec{\phi}.
    \label{eq:intensity}
\end{equation}
This is a familiar algebraic form, the Quadratic Form. Now that there is an equation relating the coupled power to the mode distribution and the electric field input, the optimization problem can be defined,
\begin{equation}
    \max_{|\vec{\phi}|^2 = 1}{I}.
\end{equation}
With this cost function, the power coupling into mode $\vec{\phi}$ is maximized under the constraint that $\phi$ is a normalized vector. This is an optimization problem that can be recognized as finding the maximum value of a Rayleigh Quotient. This is done by finding the eigenvector of $W\vec{E}\vec{E}^HW$ with the largest eigenvalue. A simple example can be seen in Figure \ref{fig:optimal_on_axis}, where the optimal mode is found for a circular aperture with a \edited{varying} central obscuration \edited{ratio}. The optimal mode is \edited{obviously} the aperture itself because the optimization problem only contained a single input mode, the on-axis wavefront. Instead of finding the eigenvector with largest eigenvalue, one can also find the solution with the Singular Value Decomposition (SVD) of $W \vec{E}$. The optimal solution is then the mode that corresponds to the largest singular value. The SVD is a more efficient approach because the dimension of the problem is smaller. Even more computational efficiency can be gained by only finding the $K$ largest singular values and modes instead of all modes\edited{, which allows fast iterative eigenvector solvers to be used \cite{arnoldi1951principle}.}

The optimization problem has to be extended to include  extended objects or systems that \edited{need} to operate in a range of conditions. If the incoming wavefronts are spatially and temporally incoherent, the total intensities add,
\begin{equation}
    I_{T} = \int_R \int_t \alpha \cdot \alpha^H dtdR.
\end{equation}
The total intensity $I_{T}$ is the intensity integrated over time $t$ and object space $R$. Substituting Equation \ref{eq:intensity} results in,
\begin{equation}
    I_{T} = \int_R \int_t \vec{\phi}^H W \vec{E}\vec{E}^H W \vec{\phi} dtdR.
\end{equation}
Both $\phi$ and $W$ can be taken out of the integral, because the fiber mode and the weight matrix do not depend on either time or the object space.
\begin{equation}
    I_{T} = \vec{\phi}^H W \int_R \int_t \vec{E}\vec{E}^H dtdR W \vec{\phi}.
\end{equation}
From this equation follows that the electric-field matrix becomes the time and object averaged covariance matrix $M$, 
\begin{equation}
    I_{T} = \vec{\phi}^H W \langle M \rangle_{t,R} W \vec{\phi}.
\end{equation}
The total intensity has been reduced to the same form as Equation \ref{eq:intensity}, and can therefore also be optimized in the same way. Here I used time and object space to average the coherence matrix, however different instrument aberrations can also be included, for example when making the system robust against tip/tilt or focus errors.

\subsection{Effects of aberrations on the coupling efficiency}
\edited{The coupling efficiency can be limited by static and random wavefront errors. To analyze the effects of both, the electric field is expanded into the diffration-limited part, the aberrated electric field and the random contribution. This leads to,}
\begin{equation}
    E = \beta E_d + \gamma E_a + \delta E_r
\end{equation}
\edited{Here $E_d$, $E_a$ and $E_r$ and the diffraction-limited part, the aberrated part and the random part, respectively. Each electric field is normalized such that the total power is 1. The coefficients $\beta$, $\gamma$ and $\delta$ are the respective contribution of each part. Both the aberrated and random part are orthogonal to the diffraction-limited field, and therefore the overlap integral between these electric fields is zero. The total intensity after integrating over space becomes,}
\begin{equation}
    I = \beta^2 + \gamma^2 + \delta^2.
\end{equation}
\edited{The ratio between the diffraction-limited part and the total intensity is the Strehl ratio, $S=\beta^2 / I$. The coupling efficiency for an arbitrary mode can be calculated with Equation 2,}
\begin{equation}
E = \beta \langle \phi E_d \rangle + \gamma \langle \phi E_a \rangle + \delta \langle \phi E_r \rangle.
\end{equation}
\edited{The angle brackets have been used to indicate the coupling integral over space. The coupled power is then,}
\begin{equation}
I = \beta^2 |\langle \phi E_d \rangle|^2 + \gamma^2 |\langle \phi E_a \rangle|^2 + \delta^2 |\langle \phi E_r \rangle| ^2 + 2\beta\gamma \Re{\{\langle \phi E_d \rangle \langle \phi E_a \rangle^\dagger\}}.
\label{eq:coupling_strehl}
\end{equation}
\edited{The cross-term between the random field and the other two fields is zero after averaging over time, which is the reason why those terms were neglected. Now, suppose that the optimal mode has been chosen before considering any static aberration, then the optimal mode is the aperture function. This means that all overlap integrals that do not contain the diffraction-limit field become zero, and the coupling efficiency becomes $\beta^2/I$ and thus equal to the Strehl of the optical system. This is not a surprise because the Strehl ratio is defined as the coupling efficiency into the on-axis Fourier mode which is equal to the aperture mode. The linear relation between the coupling efficiency and the Strehl ratio has been shown to hold on-sky in the presence of random wavefront errors \cite{faucherre2000using, jovanovic2017efficient}. However, relation between the coupling efficiency and the Strehl was found to be $\eta=a \cdot S$ with $a$ some coefficient between 0 and 1. This is a result of not matching the fundamental mode of the fiber to the aperture function. The scaling coefficient $a$ is equal to $|\langle \phi E_d \rangle|^2$, the diffraction-limited coupling efficiency. A value below 1 shows how much the used fiber mode deviates from the optimal mode in terms of throughput.}

\edited{The coupling efficiency can be improved if the static aberration is known. However, most of the current ExAO systems have slowly evolving wavefront errors that are not corrected by the AO system. These occur due to environmental variations (e.g. temperature) or even a changing gravity vector. It is not possible to correct for these wavefront errors, as they are random. The only errors that are correctable through manipulation of the fiber mode are the wavefront aberrations intrinsic to the optical design of the system. And even then, only a single field point can be corrected if there is a field dependence of the aberrations. Therefore, it is most likely the easiest to design a beam reshaper to create the aperture mode, and use focal plane wavefront sensing techniques to clean up the static aberrations \cite{jovanovic2018review}.}

\subsection{Tiling the focal plane with identical fibers}
The previous section setup the optimization problem for a single fiber. This section extends the problem to a tiled focal plane. For the tiled focal plane system it is \edited{numerically} more efficient to work in the focal plane, because the focal plane can be sampled with less pixels than the pupil plane. The optical system under consideration has a propagation matrix $C$ that relates the pupil plane electric field to the focal plane electric field. The focal plane electric field becomes,
\begin{equation}
    E_f = CE.
\end{equation}
The fiber mode in focal plane is a new mode, $\psi$. The coupling into this mode is,
\begin{equation}
    \beta = \psi^H W_f CE.
\end{equation}
Here $W_f$ is the focal plane weight matrix. This matrix can be used to select the area in the focal plane that will couple into the fiber by setting the weight to 1 when the pixel is part of the fiber mode and 0 when it is not. Adding the propagation matrix leads to the following total intensity,
\begin{equation}
    I_{T} = \vec{\psi}^H W_f C \langle M \rangle_{t,R} C^H W_f \vec{\psi}.
\end{equation}
This is still only the equation for a single fiber. To include tiling of the focal plane, the intensity has to summed over all fiber in the focal plane,
\begin{equation}
    I_{T} =\sum_n \vec{\psi}\left(\vec{r}-\vec{r}_n\right)^H W_f C \langle M \rangle_{t,R} C^H W_f \vec{\psi}\left(\vec{r}-\vec{r}_n\right).
\end{equation}
Each fiber has a different position, $\vec{r}_n$, in the focal plane. A shift in the focal plane is a tilt in the pupil plane. Instead of moving the fiber mode, the PSF can also be shifted by giving the incoming wavefront an opposite tilt in the pupil plane,
\begin{equation}
    I_{T} =\sum_n \vec{\psi}^H W_f C  D\left(\vec{r}_n\right) \langle M \rangle_{t,R} D\left(\vec{r}_n\right)^H C^H W_f \vec{\psi}.
\end{equation}
The tilt is added by the matrix $D\left(\vec{r}_n\right)$. The focal plane mode does not depend on $n$ anymore and can be taken out of the summation together with the weight matrix,
\begin{equation}
    I_{T} =\vec{\psi}^H W_f \left[\sum_n C D\left(\vec{r}_n\right) \langle M \rangle_{t,R} D\left(\vec{r}_n\right)^H C^H \right] W_f \vec{\psi}.
\end{equation}
This equation can be simplified by replacing the sum and matrix products with a single system matrix $S$,
\begin{equation}
    I_{T} =\vec{\psi}^H W_f S W_f \vec{\psi}.
\end{equation}
And now the equation is reduced to the familiar form of a Rayleigh quotient again. The optimal mode can be found by finding the eigenvector with the largest eigenvalue of $W_f S W_f$. \edited{The effects of NCPA and random aberrations is more difficult to calculate analytically for the fiber array. The effects of NCPA will, most likely, have less of an influence on the throughput for a fiber array. Any aberration can be described as a sum of Fourier modes. The Fourier modes create a satellite speckle at their spatial frequency and that speckle still couples relatively efficiently if it is within the field of view of the fiber array. However, the speckle will not couple if it is created outside the field of view. For the single object spectroscopy case, any aberration lowers the throughput which makes the analysis straightforward. While for the fiber array, the coupling efficiency depends on the spatial frequency of the aberration, the strength of the aberration (higher-order terms may go outside the field of view), the pitch of the fiber array, and the considered field of view. This makes the throughput loss very system dependent. Although as a general rule of thumb, the throughput depends linearly on the Strehl of the system for random aberrations because the total power is equal to Equation \ref{eq:coupling_strehl} summed over all fibers.}

\section{Comparison between optimal fiber modes and conventional fibers}
\subsection{Single-mode fiber spectroscopy}
\begin{figure*}[htbp]
 \centering
 \includegraphics[width=\textwidth]{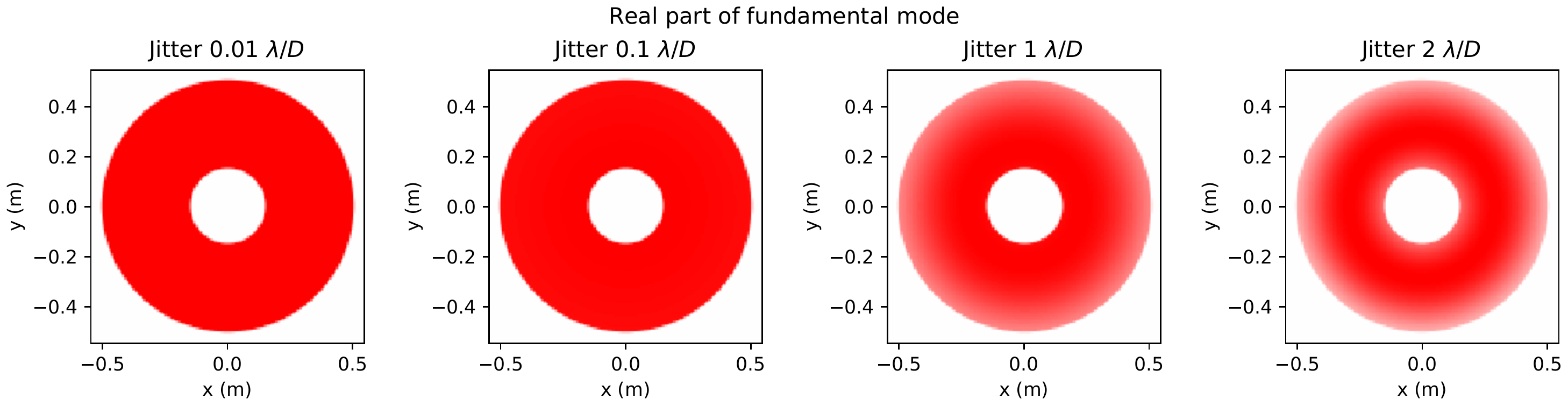}
 \caption{The optimal mode as function of jitter strength. The optimal mode for small amounts of jitter is the aperture function. If the strength is increased the edges of the mode are apodized and become softer.}
 \label{fig:jitter_modes}
\end{figure*}

For single or multi-object spectroscopy there is a single fiber per object. The optimal mode would be the aperture function itself, which has been shown in Figure \ref{fig:optimal_on_axis}. However, most telescopes have jitter of the Point Spread Function (PSF). The optimal mode in presence of jitter changes depending on the amount of jitter. Figure \ref{fig:jitter_modes} shows the optimal mode for increasing amounts of PSF jitter. Similar to the case of only an on-axis wavefront, the optimal mode for a telescope with jitter is also completely real. For small amounts of jitter, the optimal mode is \edited{the} aperture again. If the jitter is increased the edges of the mode get apodized, and the mode becomes more Gaussian like. The apodization of edges happens because tip/tilt errors cause a large differences in phase at the edges. To increase the average throughput the influence of the electric field at the edges has to be decreased with respect to the center, which is why the edges are softer for large amounts of jitter.

\begin{figure*}[htbp]
 \centering
 \includegraphics[width=\textwidth]{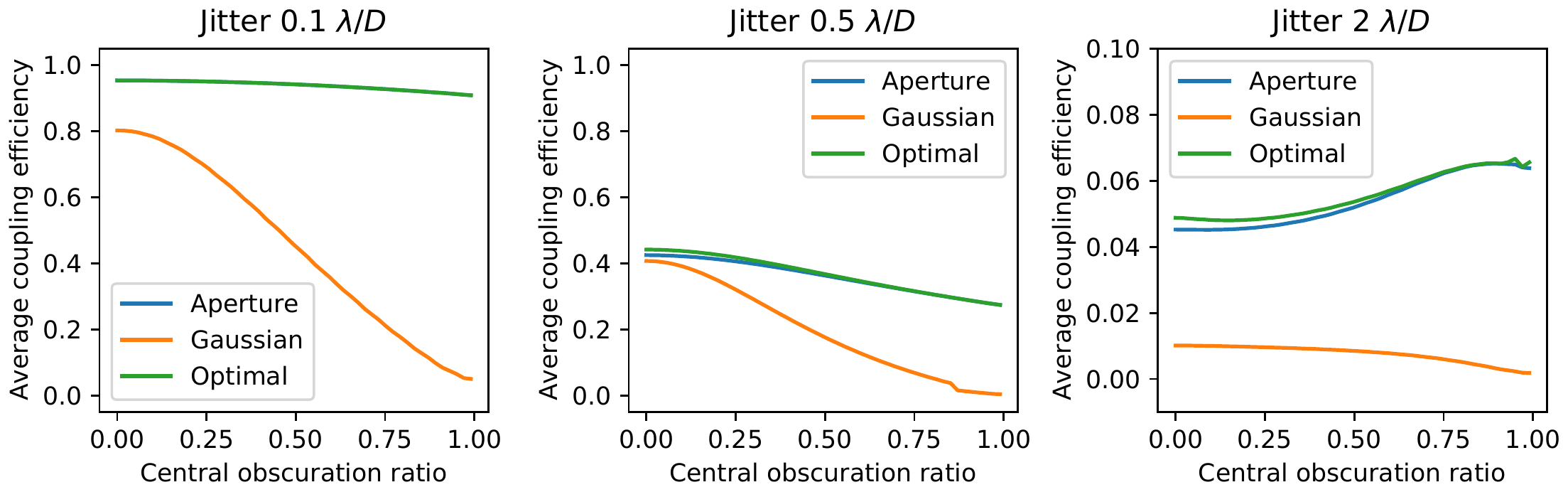}
 \caption{The average throughput as function of the relative size of the central obscuration for 3 different jitter strengths. Each color represents a different fiber mode. The blue line uses the aperture as the mode. The orange is a size optimized Gaussian, and the green is the optimal mode. The optimal mode overlaps for almost all situations with the aperture mode. For large central obscurations, the Gaussian fiber mode loses a significant amount of flux.}
 \label{fig:jitter_throughput}
\end{figure*}

The average throughput as function of central obscuration is shown in Figure \ref{fig:jitter_throughput}. I compare 3 different fiber modes. The first is the aperture mode itself, as this was shown to be the optimal mode for an unaberrated on-axis wavefront. The second is an optimized Gaussian mode, which is a good approximation to the fundamental mode of a step-index fiber. And the third mode is the optimal mode as found by Equation 10. There is only a small difference between the throughput of the aperture mode and the optimal mode. For jitter between $0.5 \lambda/D$ and $1\lambda/D$, the Gaussian mode has a throughput that is very close to the throughput of the optimal mode. The difference in throughput between the Gaussian mode and the optimal mode increases when the central obscuration is larger. The main reason for this is that the Gaussian mode has most of its power at the center, which is exactly the part where there is no light for obscured telescopes. These results demonstrate that there is no need to optimize the mode for jitter, the aperture mode itself is close to optimal for all jitter strengths. System with large amounts of residual jitter will benefit more from reducing the jitter than from trying to match the fiber mode to the aperture. The throughput gain for a central obscuration of 0.2 (VLT-like) with 0.5 $\lambda/D$ jitter is only 5 percent points, while reducing the jitter to 0.1 $\lambda/D$ will increase the throughput by a factor 2.

\subsection{Single-mode fiber integral-field spectroscopy}
Integral-field spectroscopy (IFS) is often used when the object of interest is extended, or the position of the object of interest is unknown. In both cases it is necessary to have a continuous sampling of the field of view. To sample such a field of view, the focal plane is tiled with a square geometry where each fiber is separated by a pitch $P$. To reach the highest fill factor, the size of each fiber should be equal to the pitch. Because the intensity distribution of the science objects is unknown, there should be no preference for any point in the field of view. Therefore, a uniform square distribution equal in size to the fiber source is chosen as the extended source. This effectively means that the average throughput over the field of view of a single fiber will be optimized. Only the field of view of a single fiber has to be considered, because a shift of $P$ will lead to the exact same coupling due to the periodic nature of the geometry. This will of course only hold for infinitely large arrays, however the edge effects are negligible for fiber arrays that are larger than the source.

\begin{figure*}[htbp]
 \centering
 \includegraphics[width=\textwidth]{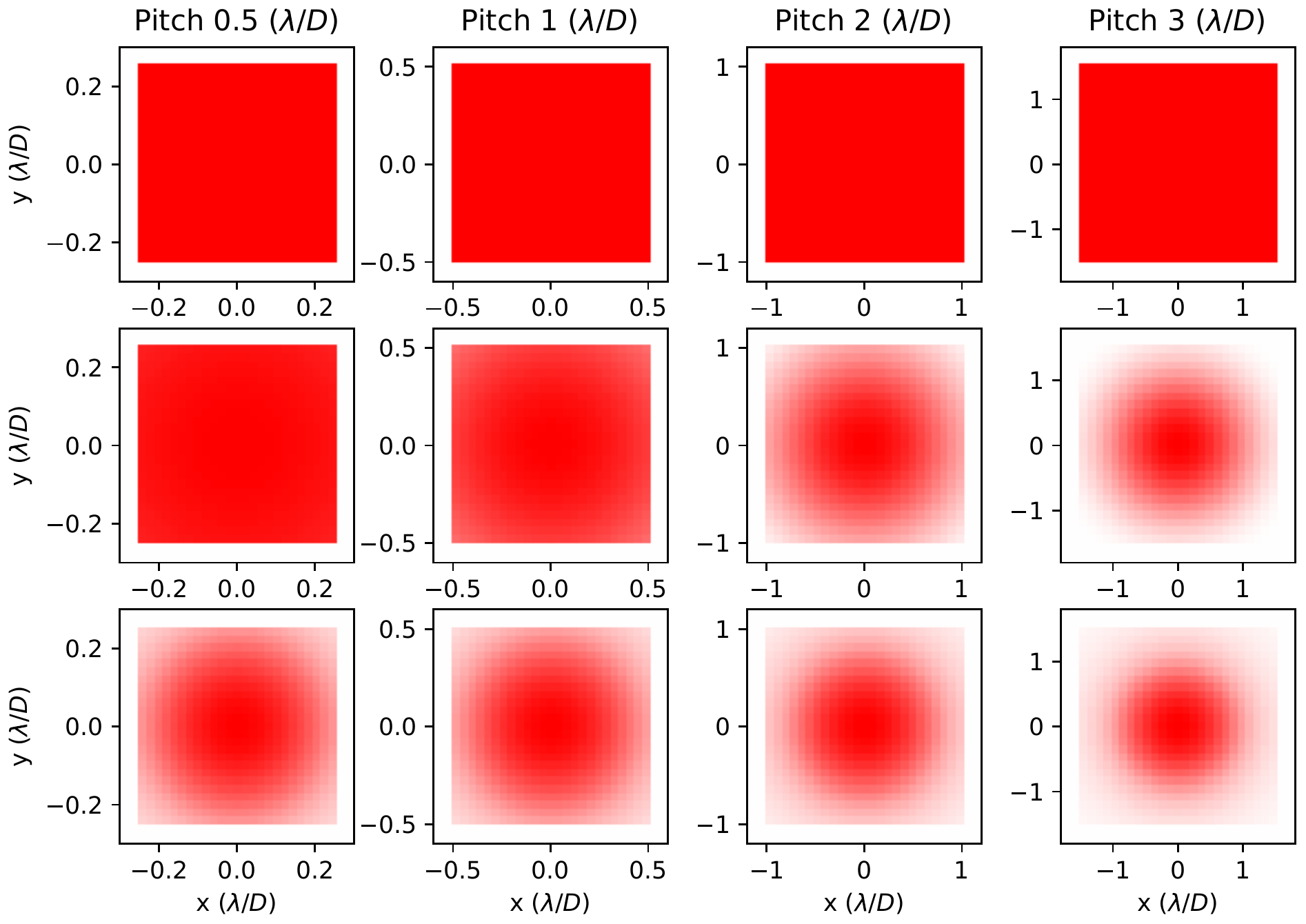}
 \caption{An overview of the modes used for the tiled focal plane array. The top row shows the uniform electric-field mode. The middle row shows the calculated optimal mode. And the bottom row shows the optimized Gaussian mode. The size of the focal plane pitch increases from from the left column to right column. The optimal mode changes from the uniform mode to the Gaussian as the pitch is increased. }
 \label{fig:optimal_tiling_mode}
\end{figure*}

Three fiber modes are considered for the tiled focal plane. For the IFS observations, the coupling should be independent of the position. A uniform square mode with the size of the fiber area is an obvious choice for such a mode. The second is a size-optimized Gaussian mode, which again represents conventional step-index fibers. And the third mode is the optimal mode as calculated by the procedure described in Section 2. An overview of the three modes for different pitch sized can be seen in Figure \ref{fig:optimal_tiling_mode}. A circular aperture was used to create the optimal modes. The field distribution of the optimal mode shows an interesting pattern. For small pitches the optimal mode is actually the uniform mode, and when the pitch increases in size the mode is apodized at the edges, and finally transforms into a Gaussian-like mode. The convergence to the uniform mode can be explained, because as the pitch becomes smaller the variation of the electric field across the fiber area also becomes smaller. In the limit of $P\to 0$, the electric-field even becomes uniform, which is why the uniform mode is the optimal mode for fiber arrays with small pitches.

\begin{figure*}[htbp]
 \centering
 \includegraphics[width=\textwidth]{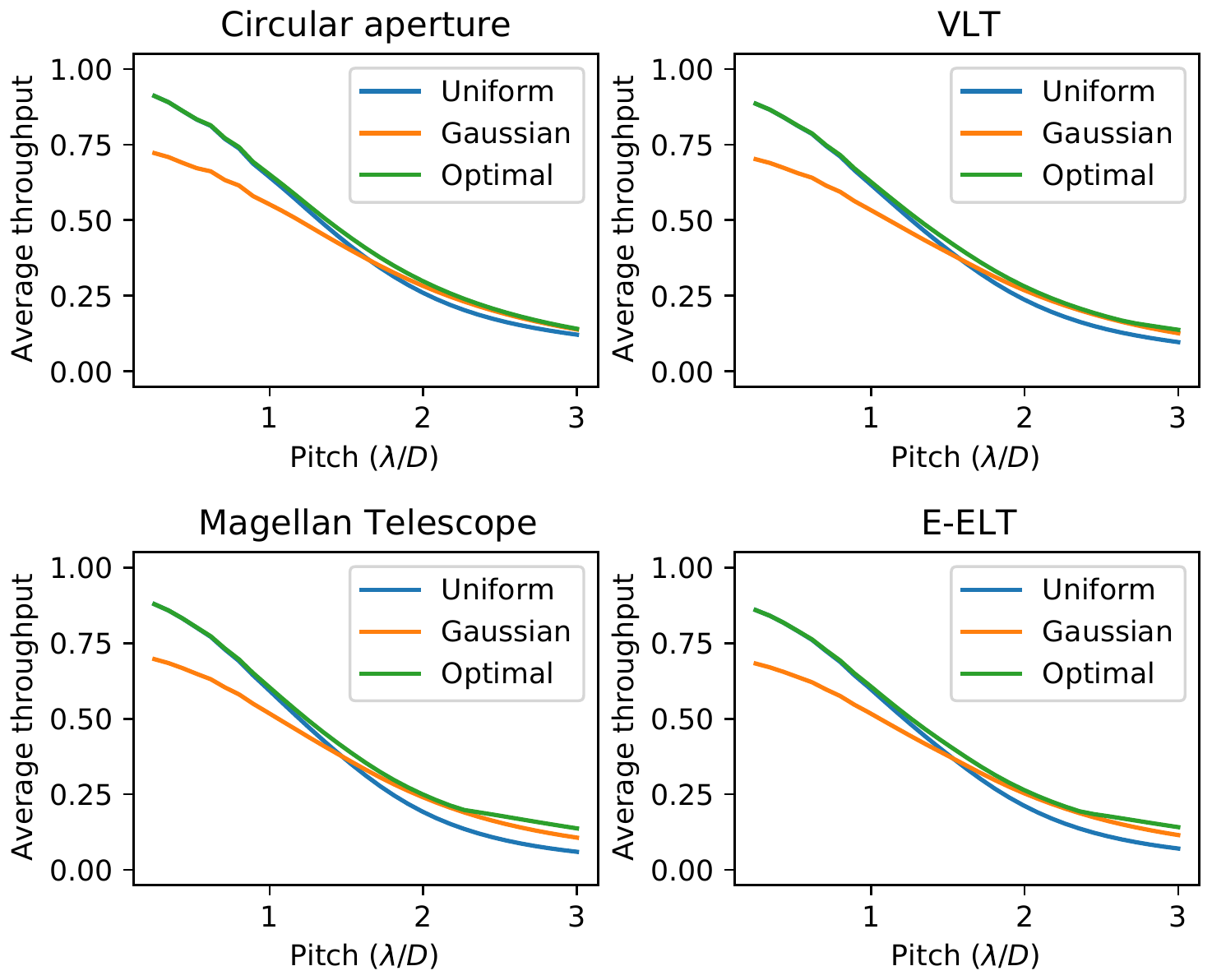}
 \caption{The spatially averaged throughput as function of pitch of the SMF. The different colors correspond to different mode field distributions. The blue lines are for a uniform mode, the orange for an optimized Gaussian and the green are the optimal modes. The average throughput is shown for 4 different telescopes, with increasing central obscuration (top left to bottom right). For small pitches the optimal mode behaves exactly like the uniform mode, while for larger pitches the performance is similar to the Gaussian mode. }
 \label{fig:fundamental_sm_throughput}
\end{figure*}

With the modes at hand, the actual average throughput over the fiber area can be found. The results of the average throughput as function of pitch for 4 different telescope apertures can be seen in Figure \ref{fig:fundamental_sm_throughput}. The throughput for a circular aperture, the VLT, the Magellan Telescope and the E-ELT are shown. This sequence of telescopes has an increasing central obscuration ratio, which is why the coupling into the Gaussian fiber modes decreases with telescope aperture. The average throughput of the optimal mode converges to the throughput of the uniform mode for small pitches and converges to the Gaussian throughput for large pitches. The transition for all apertures is around $1.5\lambda/D$. At that size, the fiber will see not only the Airy core, but also the first Airy ring which has opposite phase of the core. For the uniform mode this reduces the coupling more than for the Gaussian mode, because the Gaussian mode has more weight at the center where the core is than the edges where the Airy ring with opposite phase is. A high throughput of at least 80\,\% can be achieved with the optimal mode when the PSF is (super-) Nyquist sampled.

For all apertures and mode fields, the average throughput decreases as the pitch is increased. This is a fundamental limit of a single-mode fiber. Efficient coupling over large fields of view can only happen if all modes are the same, but the electric fields quickly decorrelate with increased tip/tilt angles. Even with the ability to choose an optimized mode, the average throughput of a single-mode fiber array with large pitches is limited.

\subsection{Few-mode fiber integral-field spectroscopy}
Few-mode fibers could be a solution for large pitches \cite{horton2007coupling}. This has been discussed in the context of step-index fibers, which support the LP modes \cite{horton2007coupling, corbett2009sampling}. A similar question can be raised, about how many modes do we fundamentally need to reach a certain throughput for a given pitch size. Previous work chose a minimal throughput of 80\,\% as the threshold of efficient coupling \cite{corbett2009sampling}, which will also be used in this work. The same optimization procedure is used for the few-mode fiber modes as was used for the single-mode case. The only difference is that the first $K$ largest eigenmodes are found, instead of only the largest.

\begin{figure*}[htbp]
 \centering
 \includegraphics[width=\textwidth]{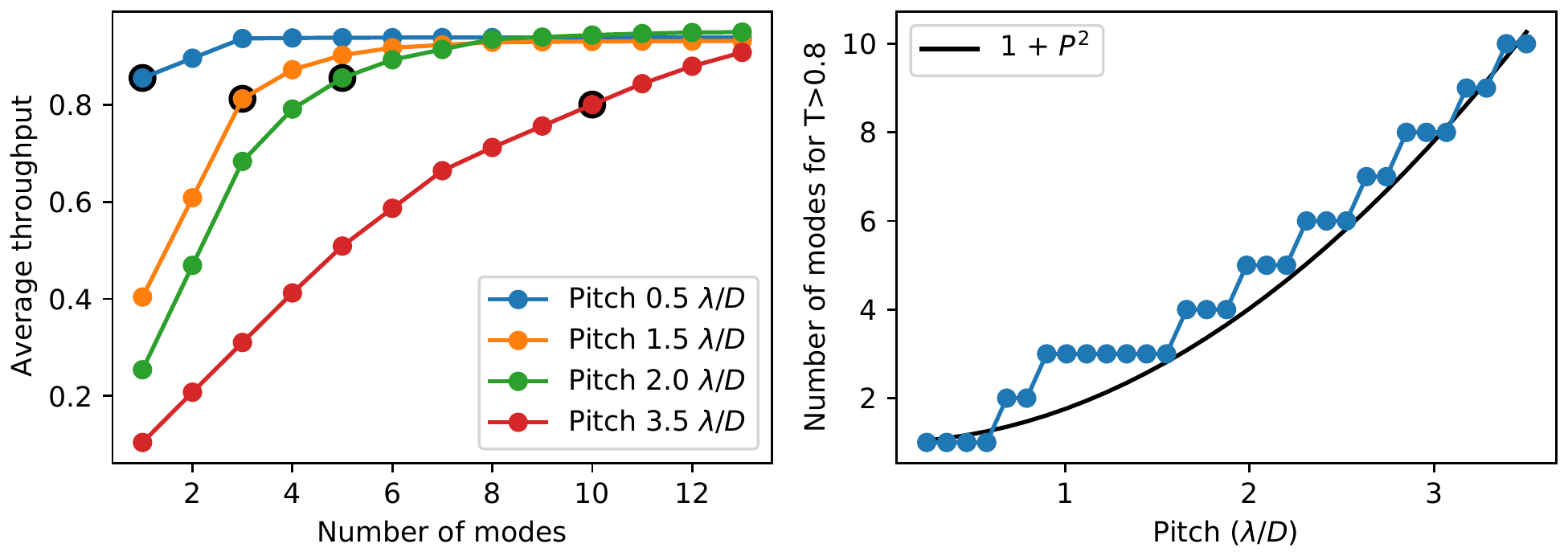}
 \caption{The figure on the left shows the throughput as function of the number of optimal modes for 4 different pitches. Each color represents a different pitch size. A black marker is used to indicate when the throughput goes above 80\,\%. The figure on the right shows the minimal number of modes that are required to achieve at least 80\,\% throughput as function of pitch size. The black line corresponds to the pitch squared to show the growth rate of the number of modes. }
 \label{fig:few_mode_limit}
\end{figure*}

The results are summarized in Figure \ref{fig:few_mode_limit}. As the pitch is increased there are more modes necessary to reach at least an 80\,\% average throughput. The number of required modes roughly scales with the square of the pitch. The number of optimal modes required is lower than the number of required step-index modes for the pitches with sizes comparable to the diffraction-limit. For a pitch between  $1.5 - 2.0 \lambda/D$, step-index fibers need at least 5 modes to reach the throughput threshold \cite{corbett2009sampling}. If an optimal set of modes could be chosen, only 3 are needed for the case of $P=1.5\lambda/D$ and 5 for $P=2.0 \lambda/D$. For larger pitches the number of modes are similar, and step-index fibers are near optimal. This means that creating an optimal mode basis for the fiber, only matters in the small pitch regime ($P<1.5\lambda/D$). The right graph of figure \ref{fig:few_mode_limit} also reveals when it is necessary to switch to few-mode fibers to maintain throughput, which happens around a pitch of $0.6 \lambda/D$. The results show that efficient single-mode integral-field spectroscopy can be done if the PSF is at least Nyquist sampled, and if the fiber mode is reshaped into the optimal mode. Another benefit is that the throughput is now jitter independent, because the throughput is not sensitive anymore to the position of the PSF. A small IFU may be preferable for some instrument if the telescopes has large residual jitter that can not be removed.

\section{Discussion and conclusion}
The fundamental throughput limit for SMF spectrographs has been investigated. The optimal mode for the single-object spectrographs is found to be the aperture mode. Even when PSF jitter is included, the aperture mode stays near optimal.

For integral-field spectrographs the optimal mode depends on the spatial sampling of the focal plane. For dense sampling, a uniform mode is optimal, while for sparse sampling, the mode of a conventional SMF is near optimal. The results also show that a high throughput (>80\,\%) can be achieved when the focal plane is (super) Nyquist sampled with optimal modes. Conventional fibers reach a throughput between 60\,\% and 65\,\% depending on the actual aperture shape, which is why single-mode tiling of the focal plane was disregarded as an option in previous work \cite{corbett2009sampling}. For the Nyquist sampled cases, the optimal mode has almost 20\,\% more throughput than a conventional SMF.

These results are important for the future success of single-mode integral-field spectrographs. At the moment, most SM IFUs for exoplanet characterization use or are planning to use relatively large pitches. These lead to regions of zero throughput \cite{por2020single, chazelas2020ristretto}, which severely limits the use of a SMFs. Two observations with different clock angles are proposed to smooth out the dead zones. This lowers the effective integration time by 2, which can also be seen as a factor 2 reduction in throughput. A pitch of $0.5 \lambda/D$ with an optimal mode already has more than 3 times the throughput of a $2.0\lambda/D$ Gaussian mode (the sampling choice for several instruments). This makes it attractive to spread the light over more SMFs at the cost of more detector noise. \edited{For a point source it may be more efficient to switch to a single-object spectrograph when the position of the point source is accurately known, because a single fiber can capture 100\% of the light if it is shaped into the aperture mode.} However, efficient single-mode integral-field spectroscopy can be done with pitch sizes up to $0.6\lambda/D$. If the pitches become larger, it may be necessary to switch to optimized few-mode fibers to retain throughput. And for fibers larger than $2\lambda/D$, the optimal mode set is very close to the modes of conventional step-index fibers and there is no substantial throughput to gain anymore.

A next step in this work will be to design and manufacture an aspheric multi-element array and test whether the step-index mode can indeed be reshaped into the optimal mode pattern.

\section{Acknowledgements}
Support for this work was provided by NASA through the NASA Hubble Fellowship grant \#HST-HF2-51436.001-A awarded by the Space Telescope Science Institute, which is operated by the Association of Universities for Research in Astronomy, Incorporated, under NASA contract NAS5-26555. This research made use of HCIPy, an open-source object-oriented framework written in Python for performing end-to-end simulations of high-contrast imaging instruments \cite{por2018high}.

\section{Disclosures}
The authors declare no conflicts of interest.

\bibliography{references}

\end{document}